\documentclass[12pt]{revtex4}
\usepackage{pslatex}
\usepackage[T1]{fontenc}
\usepackage{epsf}

\begin{document}
\title{Explicitly correlated wavefunctions of the ground state and the lowest
quintuplet state of the carbon atom}
\author{Krzysztof Strasburger}
\affiliation{Department of Physical and Quantum Chemistry \\
Faculty of Chemistry \\
Wroc{\l}aw University of Science and Technology \\
Wybrze\.{z}e Wyspia\'{n}skiego 27, 50-370 Wroc{\l}aw, Poland}
\date{\today}

\begin{abstract}
Variational, nonrelativisitic energies
have been calculated for the ground state ($^3P_g$) and the lowest quintuplet
state ($^5S_u$) of the carbon atom, with wavefunctions expressed in the
basis of symmetry-projected, explicitly correlated Gaussian (ECG) lobe
functions. New exact limits of these energies have been estimated,
amounting to $-37.844906(4)$ and $-37.691751(2)$ hartree. With finite
nuclear masses and leading, scalar relativistic corrections included,
respective experimental excitation energy of $^{12}C$
has been reproduced with accuracy of about 7 cm$^{-1}$.
\end{abstract}

\maketitle

\section{Introduction}

Carbon has the richest chemistry among all elements. Literally thousands of
scientific papers, reporting various calculations on carbon compounds, are
written every year. On the other hand, determination of the carbon atom
properties on the grounds of theory, with accuracy comparable to that
offered by spectroscopic experiments, remains a challenge for
computational chemistry. The present work does not provide an ultimate
solution of this problem. The aim is more modest -- to demonstrate that 
an explicitly correlated wavefunction is able to yield about
10~$\mu$hartree energy accuracy for six-electron atom and that an ansatz
with inaccurate angular dependency of basis functions may prove to be more
efficient from that which is seemingly better or at least more elegant.

For a long time, the best variational nonrelativistic energy of the carbon
atom ground state was that obtained with the configuration interaction method
in the year 1974 \cite{C-Sasaki}, using Slater orbitals corresponding to
the angular momentum quantum number $l$ up to 6, and up to quadruply excited
configurations. That result still represents the most
accurate published CI energy of this state. Recent CI calculation
in which configurations were selected carefully, considering their energy
contributions, but built from orbitals the with $l$ limited to 4, gave a
little higher energy \cite{C-Ruiz}. Comparison with the work devoted to boron
atom and anion (the latter being isoelectronic with carbon atom's ground
state) \cite{B-Almora-Diaz},
makes it clear that orbitals with much higher $l$ are needed for building
a many-electron basis, capable of yielding the accuracy of about 1 mhartree.
CI variational energies have been surpassed by these obtained in calculations
with explicitly
correlated Gaussian functions \cite{C-Sharkey, ChemRev}, but the best result
reported is still about 1.5~mhartree above the estimate of exact
nonrelativistic energy \cite{Chakravorty}. This estimate is approached well by
nonvariational (or at least not strictly variational) methods -- coupled
clusters with exponential correlation factor (CC-F12) \cite{Noga-F12},
diffusion quantum Monte Carlo
simulations \cite{C-Maldonado,atoms-Seth} and the ``free complement''
method \cite{FC-Nakatsuji}. In the latter, regularized Krylov
sequences of functions generated by the system's Hamiltonian are used as the
basis. The results, reported in cited references, differ however even by few
milihartree and those, for which standard deviations are
given \cite{atoms-Seth,C-Maldonado,FC-Nakatsuji}, do not overlap (table
\ref{encomp} in further text).
Calculations on the $^5S_u$ state (the one spin quintuplet below the
ionization limit of the carbon atom, known widely for the $sp^3$ orbital
hybridization model) \cite{C-Sasaki,C-Maldonado,FC-Nakatsuji} yielded
somewhat more consistent results, with the discrepancies reaching
several hundreds microhartrees. 

A comparison of theoretical results with spectroscopic data does not require
absolute energies of states. Good agreement with experimental excitation energy
of the carbon atom from the ground state to the $^5S_u$ state (and other
states too, but they are not the subject of present work) has been achieved
in multiconfiguration Hartree-Fock calculations \cite{C-MCHF}, with
omitted correlation of $1s$ core electrons. According to the same reference
publication, the leading relativistic energy corrections contribute about
90 cm$^{-1}$. They are partially taken into consideration in the present work.

For few-electron systems, for which high accuracy is desired,
an explicitly correlated ansatz, used in the variational framework, is
most effective. Unfortunately, the associated computational cost grows rapidly
with the number of electrons. Despite of technological progress, both on
computer hardware and software side, almost two decades passed between first
publications of the ground state energies, computed with nearly-microhartree
accuracy, for beryllium \cite{Be-KCR} and boron \cite{B-BA,B-PKP} atoms.
Concerning the analytical forms of explicitly correlated basis functions,
many of them are tractable for two- and three-electron systems, while only
the Hylleraas-CI method and ECGs are competitive in practice for four-electron
atoms and atomic ions. Both methods
provided energies accurate to few nanohartrees \cite{Be-Sims,Li-Sims,Be-PPK}.
The latter ansatz is at present the one applicable for systems containing
five and more electrons, due to relatively simple form of integrals appearing
in Hamiltonian matrix elements. Other types of basis functions are used
too, but occurence of complicated many-electron integrals, without known
analytical solutions, forces resorting to stochastic techniques
\cite{atoms-Seth,C-Maldonado,FC-Nakatsuji} or using the resolution
of identity \cite{Noga-F12} for reduction of their complexity.

In the present calculation, the ansatz of explicitly correlated Gaussian
lobe functions (called also ``Gaussians with shifted centers'',
see Ref. \cite{MVD} for an example) is employed.
This ansatz was applied, with a success, in studies
of small molecules, molecular ions and van der Waals complexes
\cite{H2H3+KCR,HeHeKR1,HeHeKR2,H3+PA,LiH-Tung}.
Free atoms have spherical symmetry, therefore their exact wavefunctions are
eigenfunctions of not only Hamiltonian, but also square of angular momentum
($\hat{L}^2$) and $z$-component of angular momentum ($\hat{L}_z$) operators.
Basis functions $\chi$ for atomic states are constructed, as a rule, so that
the relations $\hat{L}^2\chi=L(L+1)\chi$ and $\hat{L}_z\chi=M\chi$ are
fulfilled {\em a priori} for particular values of $L$ and $M$ quantum
numbers \cite{B-BA,B-PKP,C-Sharkey,Be-P-Bubin,Li-D-Bubin,L3-Sharkey, N-Sharkey}.
On the contrary, a lobe function centered off the nucleus is not an
eigenfunction of these operators. Convergence towards desired state may
however be enforced by variational optimization of trial wavefunction, with
proper symmetry constraint. This method, introduced in
earlier papers devoted to high $L$ states of the lithium atom \cite{Li-KS}
and various states of many-electron harmonium
\cite{harm3-JC,harm4-JC,harm4W-JC,harm56-JC,harm6-KS},
will be shortly described in next section.
Atomic units are used unless stated otherwise.

\section{Methods}

\subsection{Nonrelativistic wavefunction}

The stationary Schr\"{o}dinger equation is solved with the nonrelativistic
Hamiltonian of n-electron atom
\begin{equation}
\hat{H}=-\frac{\nabla^2_{nuc}}{2m_{nuc}}+\sum_{i=1}^{n}\left(-\frac{\nabla^2_i}{2}
-\frac{Z}{r_i}\right)+\sum_{i>j=1}^{n}\frac{1}{r_{ij}}
\label{hamilt}
\end{equation}

The wavefunction, depending on spatial (${\bf r}_i$) and spin ($s_i$)
coordinates, 
\begin{equation}
\Psi({\bf r}_1,s_1,\ldots,{\bf r}_n,s_n)=\sum_{I=1}^K C_I \hat{A}\Theta_I(s_1,\ldots,s_n)\hat{P}\chi_I({\bf r}_1,\ldots,{\bf r}_n)
\label{wavefn}
\end{equation}
with proper permutational symmetry ensured by $\hat{A}$
and primitives $\chi_I$ being explicitly correlated Gaussian lobe functions
\begin{equation}
\chi_I({\bf r}_1,\ldots,{\bf r}_n)=\exp{\left[-\sum_{i=1}^n\alpha_{I,i}({\bf r}_i-{\bf R}_{I,i})^2-\sum_{i>j=1}^n\beta_{I,ij}r^2_{ij}\right]}
\label{primit}
\end{equation}
is not an eigenfunction of $\hat{L^2}$ for non-zero ${\bf R}_{I,i}$ vectors.
Deviation of $\langle L^2\rangle$ from exact
$L(L+1)$ eigenvalue is effectively diminished by the procedure of
variational energy minimization, in which nonlinear parameters
$\alpha_{I,i}$, $\beta_{I,ij}$ and ${\bf R}_{I,i}$) are optimized. Linear
coefficients are determined by solution of the eigenvalue problem, for given
set of nonlinear parameters. The convergence towards
desired state is ensured by the spatial symmetry projector $\hat{P}$, proper
for an irreducible representation of selected finite point group. Action of
$\hat{P}$ upon basis functions annihilates a finite subset of their unwanted
 components whose symmetry properties are specific to others, than the desired
one, representations of the $K_h$ infinite point group. Particularly,
$A_2$ representation of the $C_{4v}$ point group was used for
the symmetry projector of the $^3P_g$ state
\begin{equation}
\hat{P}(^3P_g)=\hat{E}+\hat{C}^1_4+\hat{C}_2+\hat{C}^3_4-\hat{\sigma}_{v1}-\hat{\sigma}_{v2}-\hat{\sigma}_{d1}-\hat{\sigma}_{d2}
\end{equation}
Confinement of all ${\bf R}_{I,i}$ vectors to the $xy$ plane ensures
proper parity (even) of the wavefunction. Lifting this constraint while using
the projector proper to the $A_{2g}$ representation of the $D_{4h}$ group
offered
only negligible energy lowering at substantial increase of computation time
even for small basis sets, therefore this alternative path has been abandoned
at early stage of the work.
$A_{1u}$ representation of the $O_h$ point group was employed for the $^5S_u$
state. The symmetry projector is simply too long (48 operations) to be
written here explicitly. Identity and all rotation operators, that form the
$O$ group, enter this projector with coefficients equal to $1$, and remaining
operators (products of the former with the inversion operator) -- with
coefficients equal to $-1$.

Single spin functions:
\begin{equation}
\Theta(s_1,\ldots,s_6)=[\alpha(1)\beta(2)-\beta(1)\alpha(2)][\alpha(3)\beta(4)-\beta(3)\alpha(4)]\alpha(5)\alpha(6)
\end{equation}
for the triplet, and
\begin{equation}
\Theta(s_1,\ldots,s_6)=[\alpha(1)\beta(2)-\beta(1)\alpha(2)]\alpha(3)\alpha(4)\alpha(5)\alpha(6)
\end{equation}
for the quintuplet, are sufficient to ensure convergence to correct variational
limits, as the spatial functions are nonorthogonal.

The optimizations of basis set parameters were
carried out for infinite nuclear mass. The eigenvalue problem was then
solved, in the same basis, for various isotopes of carbon. The center of
mass (CM) motion was not separated explicitly from the Hamiltonian, as the
wavefunction depends on relative coordinates only (${\bf r}_i$ in all
equations is the vector of coordinates of $i^{th}$ electron relative to
the nucleus). In such case, total kinetic energy operator in the laboratory
coordinate frame, acting upon the wavefunction (or a basis function) gives
the same result as action of the kinetic energy operator of relative motion,
because $\hat{T}=\hat{T}_{CM}+\hat{T}_{rel}$, and $\hat{T}_{CM}\Psi_{rel}=0$
for any definition of internal coordinates \cite{BHA-Kutz}.

\subsection{Relativistic corrections}

The relativistic corrections to the energy are obtained in the perturbative
series in the fine structure constant
$\alpha=\frac{1}{4\pi\epsilon_0}\frac{e^2}{\hbar c}$.
In atomic units, the value of $\alpha$ is equal to the reciprocal
of the speed of light in vacuum, $c=137.036$.
With the rest mass energy omitted, successive terms in the expansion
\begin{equation}
E=E^{(0)}+E^{(2)}+E^{(3)}+\cdots
\end{equation}
are calculated as expectation values of respective operators, with known
nonrelativistic wavefunction. $E^{(0)}$ is the
nonrelativistic energy, $E^{(2)}$ contains the Breit-Pauli corrections and
higher order terms are the QED (radiative) corrections. The Breit-Pauli
Hamiltonian contains the $\hat{H}_{RS}$ operator, which is responsible
for the scalar relativistic correction, shifting the energies of whole
terms, and the fine and hyperfine structure operators. Only
the former is being considered in this work.
For fixed nucleus, the relativistic shift Hamiltonian
\begin{equation}
\hat{H}_{RS}=\hat{H}_1+\hat{H}_2+\hat{H}_3+\hat{H}_4
\end{equation}
consists of following components:
\begin{equation}
\hat{H}_1=-\frac{1}{8c^2}\sum_{i=1}^n\nabla^4_i
\end{equation}
is the mass-velocity correction,
\begin{equation}
\hat{H}_2=\frac{Z\pi}{2c^2}\sum_{i=1}^n\delta({\bf r}_i)
\end{equation}
is the electron-nucleus Darwin term,
\begin{equation}
\hat{H}_3=\frac{\pi}{c^2}\sum_{i>j=1}^n\delta({\bf r}_{ij})
\end{equation}
represents the sum of electron-electron Darwin term and spin-spin Fermi
contact interaction (the latter after integration over spin variables
\cite{Davidson-CH2}), and
\begin{equation}
\hat{H}_4=\frac{1}{2c^2}\sum_{i>j=1}^n\left(\frac{\nabla_i\cdot\nabla_j}{r_{ij}}+
\frac{{\bf r}_{ij}\cdot[({\bf r}_{ij}\cdot\nabla_i)\nabla_j]}{r_{ij}^3}\right)
\end{equation}
describes the interaction of magnetic dipoles arising from orbital motion
of the electrons. Expectation values of $\hat{H}_1$, $\hat{H}_2$ and $\hat{H}_3$
are known to converge slowly, because these operators sample the wavefunction
for short interparticle distances, where ECG functions have an incorrect
analytical behaviour (do not describe the wavefunction's cusps). This deficiency
may be overcome by regularization of the problem \cite{Drachman}, respective
technique is however not implemented yet in author's program, so direct
formulas were used in the calculations.

\section{Results and discussion}

The most time-consuming part of the calculations was the optimization of
nonlinear parameters of basis functions. It was commenced with very small
sets, with sizes of 1, 2 and 3 (stages 1, 2 and 3) respectively.
Basis enlarging was performed, with reuse of previously optimized functions
in mind. Therefore functions from stage $k-2$ were appended to the set
obtained at stage $k$. Consequently, the successive basis sizes
formed the Narayana's cows sequence \cite{integers}. Each new basis was
optimized, function by function, in cycles.
Then the expectation value of the $\hat{L}^2$ operator and the virial ratio of
potential and kinetic energies ($-\frac{V}{T}$) were computed. The results are
given in table \ref{table-energy}, beginning with 88 ECGs.
The convergence of the energy and $\langle L^2\rangle$ was substantially better
for the $^5S_u$ state, therefore the calculations for this state were
finished with 4023 basis functions, while basis of 5896
functions was additionally built for the ground state.

\begin{table}
\caption{Nonrelativistic energies, squares of angular momentum and
virial ratios for $^{\infty}C$. For extrapolated (E) results, standard deviations
of the least significant digit are given in parentheses \label{table-energy}}
\begin{tabular}{|c|c|c|c|c|c|c|} \hline
  & \multicolumn{3}{c|}{$^3P_g$} &  \multicolumn{3}{c|}{$^5S_u$}  \\ \hline
K & Energy & $\langle L^2\rangle$ & $-\frac{V}{T}$ & Energy & $\langle L^2\rangle$ & $-\frac{V}{T}$ \\ \hline
  88 & $-37.833555219$ & 2.00015841 & 2.000005668 & $-37.688655780$ & 0.00018743 & 2.000004222 \\
 129 & $-37.838026763$ & 2.00009725 & 1.999990991 & $-37.690127107$ & 0.00008282 & 2.000009232 \\
 189 & $-37.840615636$ & 2.00008198 & 2.000002689 & $-37.690927579$ & 0.00006823 & 2.000001843 \\
 277 & $-37.842526491$ & 2.00005762 & 1.999986861 & $-37.691351361$ & 0.00004046 & 1.999999213 \\
 406 & $-37.843672608$ & 2.00003558 & 1.999993489 & $-37.691561153$ & 0.00002250 & 2.000000133 \\
 595 & $-37.844247698$ & 2.00002376 & 1.999995273 & $-37.691666242$ & 0.00000907 & 2.000000081 \\
 872 & $-37.844533602$ & 2.00001554 & 1.999997023 & $-37.691707096$ & 0.00000554 & 1.999999854 \\
1278 & $-37.844711824$ & 2.00000879 & 1.999999330 & $-37.691727292$ & 0.00000336 & 1.999999806 \\
1873 & $-37.844794050$ & 2.00000523 & 1.999999866 & $-37.691739801$ & 0.00000138 & 2.000000038 \\
2745 & $-37.844851675$ & 2.00000255 & 2.000000012 & $-37.691745022$ & 0.00000069 & 1.999999997 \\
4023 & $-37.844877180$ & 2.00000125 & 2.000000010 & $-37.691747780$ & 0.00000036 & 1.999999992 \\
5896 & $-37.844889402$ & 2.00000071 & 2.000000040 &               &            &             \\ \hline
E    & $-37.844906(4)$ & 2          &             & $-37.691751(2)$ & 0          &           \\ \hline
\end{tabular}
\end{table}

\begin{figure}
\caption{Energy dependency on the deviation of $\langle L^2\rangle$ from
exact value \label{fig-el2}}
\epsfxsize=7.5cm \epsfbox{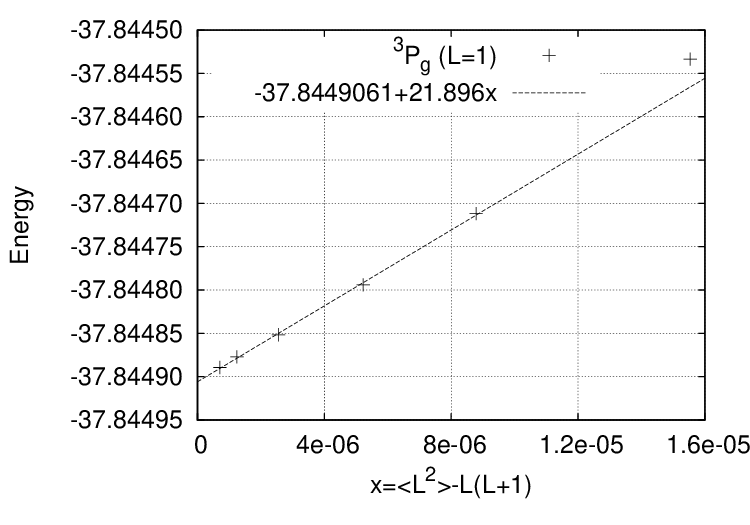} \hspace*{1cm} \epsfxsize=7.5cm \epsfbox{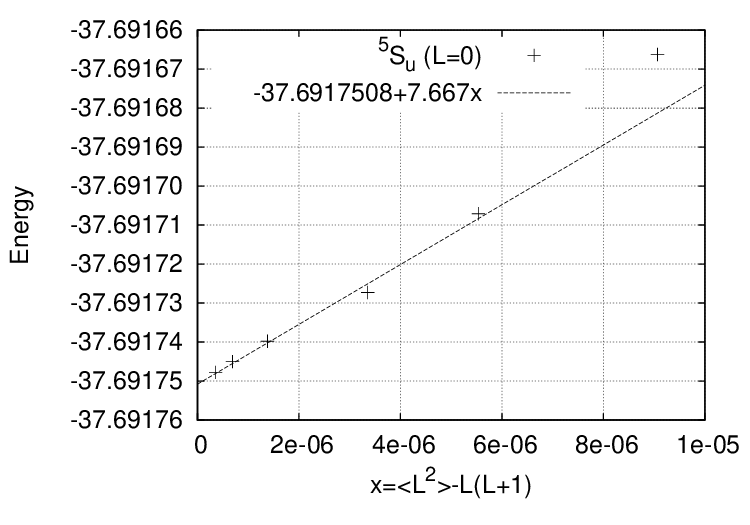}
\end{figure}

There is no regularity to be found among virial ratios. Their values, close
to 2, say only that all parameters were optimized reasonably well. No parameter
scaling based on virial ratios was attempted. On the other hand, the squares
of angular momentum converge to known exact limits, making it possible to
try to extrapolate the energies as functions of $x=\langle L^2\rangle-L(L+1)$.
There is no theoretical foundation for this extrapolation, other than an
observation, that the deviation of $\langle L^2\rangle$ is linearly proportional to the
rotation energy error, and assumption that the latter is a slowly varying
fraction of the total energy error. Extrapolation to $x=0$ with least
squares linear regression (fig. \ref{fig-el2}), using 5 points for each state,
yields estimates of exact nonrelativistic energies of the $^{\infty}C$ atom.
For the ground state, this estimate differs by about 0.1 mhartree from
the previous one \cite{Chakravorty} and is certainly more precise.

The energy convergence with projected ECG lobe functions appears better
than with ECGs mutiplied by a proper polynomial of electrons coordinates,
which are eigenfunctions of $\hat{L}^2$ (table \ref{encomp}). The energies
obtained in the present work, for 189 and 406 functions, are lower
than those published in refs. \cite{C-Sharkey,ChemRev}, computed with 500
and 1000 basis functions respectively. The best published result of
Diffusion Quantum Monte Carlo simulation \cite{atoms-Seth} is surpassed
with 872 functions, while final variational energy is lower by about
400 $\mu$hartree. This difference exceeds by far the numerical uncertainty
of that simulation. The present result is matched only by the FC-CFT
method by Nakatsuji {\em et al.} \cite{FC-Nakatsuji}, at least for the ground
state, but with significantly wider error margins.

\begin{table}
\caption{Comparison of calculated, nonrelativistic energies of $^{\infty}C$
\label{encomp}}
\begin{tabular}{|l|l|l|} \hline
method & $^3P_g$ & $^5S_u$ \\ \hline
CI-SDTQ ($l_{max}=6$)$^a$ & $-37.8393$ & $-37.6893$ \\
CI (selected configurations, $l_{max}=4$)$^b$ & $-37.8352652$ & \\
ECG, K=500$^c$ & $-37.84012879$ & \\
ECG, K=1000$^d$ & $-37.843333$ & \\
CCSD(T)-F12$^e$ & $-37.844334$ & \\
DMC$^f$ & $-37.84185(5)$ & $-37.69026(3)$ \\
DMC$^g$ & $-37.84446(6)$ & \\
FC-CFT$^h$ & $-37.845004(282)$ & $-37.690998(99)$ \\
estimated exact$^i$ & $-37.8450$ & \\
present work (ECG lobes): & & \\
variational & $-37.844889402$ & $-37.691747780$ \\
extrapolated & $-37.844906(4)$ & $-37.691751(2)$ \\ \hline
\multicolumn{3}{l}{$^a$Ref. \cite{C-Sasaki}, $^b$Ref. \cite{C-Ruiz},
$^c$Ref. \cite{C-Sharkey}, $^d$Ref. \cite{ChemRev}, $^e$ Ref. \cite{Noga-F12},
$^f$ Ref. \cite{C-Maldonado}, $^g$ Ref. \cite{atoms-Seth},
$^h$ Ref. \cite{FC-Nakatsuji}, $^i$ Ref. \cite{Chakravorty}} \\
\end{tabular}
\end{table}

Nonadiabatic calculations were carried out with nuclear masses
$m_{^{12}C}=21868.6618$,
$m_{^{13}C}=23697.6661$, and $m_{^{14}C}=25520.3468$,
calculated from known molar masses of carbon isotopes (12$u$,
13.003355$u$ and 14.003241$u$ respectively) -- dividing them by the Avogadro
number and subtracting 6 electron masses.
The same basis sets were used as for fixed nucleus, only the linear coefficients
in the wavefunctions (Eq. \ref{wavefn}) were obtained independently for each
isotope. Extrapolations were also based on an assumption that the gap from
the best variational energy to the limit, and standard deviation of
extrapolated energy, do not change with the nuclear mass.
For a given state, the differences of nonrelativistic energies
of isotopes converge very quickly and remain stable, therefore only the
results obtained with two
largest basis sets are given in table \ref{tab-isotop}. The effect of finite
nuclear mass contributes $-39.78\mu$hartree (or $-8.731$cm$^{-1}$) to the
energy of $^3P_g\rightarrow^5S_u$ excitation of $^{12}C$. The isotopic
shift between $^{13}C$ and $^{12}C$ agrees perfectly with experimental
data \cite{experiment} so a reliable prediction for the $^{14}C$ isotope
is possible with nonrelativistic wavefunctions (bottom of table \ref{tab-isotop}).

\begin{table}
\caption{Nonrelativistic energies of carbon isotopes (E for extrapolated
values); energy differences are in mhartree
\label{tab-isotop}}
\begin{tabular}{|c|c|c|c|c|c|} \hline
K & $E(^{12}C)$ & $E(^{13}C)$ & $E(^{14}C)$ & $E(^{13}C)-E(^{12}C)$ &
 $E(^{14}C)-E(^{12}C)$ \\ \hline
\multicolumn{6}{|l|}{$^3P_g$} \\ \hline
4023 & $-37.843165183$ & $-37.843297308$ & $-37.843410138$ & $-0.132125$ & $-0.244955$ \\
5896 & $-37.843177408$ & $-37.843309534$ & $-37.843422363$ & $-0.132126$ & $-0.244955$ \\
E & $-37.843194(4)$ & $-37.843326(4)$ & $-37.843439(4)$ & & \\ \hline
\multicolumn{6}{|l|}{$^5S_u$} \\ \hline
2745 & $-37.690072810$ & $-37.690201865$ & $-37.690312073$ & $-0.129055$ & $-0.239263$ \\
4023 & $-37.690075568$ & $-37.690204623$ & $-37.690314831$ & $-0.129055$ & $-0.239263$ \\
E & -37.690079(2) & -37.690208(2) & -37.690318(2) & & \\ \hline
\multicolumn{6}{l}{Isotopic shift for $D=E(^5S_u)-E(^3P_g)$} \\
\multicolumn{6}{l}{$D(^{13}C)-D(^{12}C)=0.003071$ (0.674cm$^{-1}$), experiment (Ref. \cite{experiment}): 0.670(5)cm$^{-1}$} \\
\multicolumn{6}{l}{$D(^{14}C)-D(^{12}C)=0.005692$ (1.249cm$^{-1}$)} \\
\end{tabular}
\end{table}

Even with finite nuclear mass taken into account, the nonrelativistic theory is
not sufficient to calculate accurate energy differences between atomic
states. The ground state term has a fine structure. According to
spectroscopic data \cite{experiment}, the terms of $^{12}C$, with $J=1$
and $J=2$, appear respectively at 16.4167(13)cm$^{-1}$ and 43.4135(13)cm$^{-1}$
above that with $J=0$.
The calculation of this split could not be completed
in this work, because of lacking implementation of expectation values of
spin-orbit and spin-spin coupling operators. At this stage, it is only possible to refer
theoretical results to weighted average energy of the $^3P_g$ term:
\begin{equation}
E_{exp}(^3P_g)=\frac{1}{9}\left[E(^3P_0)+3E(^3P_1)+5E(^3P_2)\right]=29.591cm^{-1}.
\end{equation}
The $^5S_u$ term appears at 33735.121(18)cm$^{-1}$, so the reference
``excitation energy'' amounts to 33705.530cm$^{-1}$.
Subtraction of extrapolated, nonrelativistic energies (table
\ref{tab-isotop}) gives 33605(1)cm$^{-1}$, which misses the
experimental result by about 100cm$^{-1}$.

The convergence of relativistic shifts and their components, for
fixed nucleus, is illustrated
with the data presented in table \ref{tab-relconv}. As expected, the
mass-velocity and electron-nucleus Darwin terms are dominant and their
convergence with increasing number of basis functions is unsatisfactory.
The differences between the results obtained in two successive largest basis
sets
still exceed 10$\mu$hartree, for both states. The expectation value of
$\hat{H}_3$, which contains two-electron Dirac delta operator, converges
also slowly but the differences fall below 1 $\mu$hartree. The orbit-orbit
interaction energies
are converged within few nanohartree. Fortunately the errors of individual
components cancel to some extent, owing to the optimization of nonlinear
parameters of the wavefunction \cite{Cencek-relat}, so that five decimal
digits of total relativistic scalar corrections for both states seem to be
stable and converged even better than their nonrelativistic energies.
There is however no perspective to extrapolate these corrections to
infinite basis set limit. The results agree fairly well with
published relativistic corrections, obtained with less accurate wave functions
\cite{Davidson-CH2,C-MCHF} ($\langle\hat{H}_1\rangle$,
$\langle\hat{H}_2\rangle$ and $\langle\hat{H}_3\rangle$). The orbit-orbit
term has been calculated only for the ground state, in the Quantum Monte Carlo
simulation \cite{atoms-Seth}.

Assuming the same scalar relativistic corrections for the $^{12}C$ isotope
as for $^{\infty}C$ (which is expected to be correct within a fraction
of $\mu$hartree for total correction \cite{Stanke-FNM}) and adding their values
obtained in the largest basis sets to extrapolated nonrelativistic energies from table
\ref{tab-isotop}, corrected energies are obtained:
$E(^3P_g)=-37.857269$ and $E(^5S_u)=-37.703662$. Their difference amounts to
$0.153607$ hartree, or 33713~cm$^{-1}$. It is not possible to calculate its
standard deviation, because of lacking error estimation for
relativistic corrections. Assuming arbitrarily that the error range is doubled,
it would amount to 2~cm$^{-1}$. The missing contribution of at least
6~cm$^{-1}$ to the excitation energy might stem from radiative corrections.

\begin{table}
\caption{Convergence of the components of relativistic shift for
$^{\infty}C$ and comparison with literature data \label{tab-relconv}}
\begin{tabular}{|c|c|c|c|c|c|} \hline
K & $\langle\hat{H}_1\rangle$ & $\langle\hat{H}_2\rangle$ &
$\langle\hat{H}_3\rangle$ & $\langle\hat{H}_4\rangle$ &
$\langle\hat{H}_{RS}\rangle$ \\ \hline
\multicolumn{6}{|l|}{$^3P_g$} \\ \hline
  88 & $-0.077592107$ & $0.062485440$ & $0.001154713$ & $-0.000016296$ & $-0.013968250$ \\
 129 & $-0.077878128$ & $0.062750198$ & $0.001139876$ & $-0.000015900$ & $-0.014003954$ \\
 189 & $-0.078232316$ & $0.063087664$ & $0.001130461$ & $-0.000015467$ & $-0.014029658$ \\
 277 & $-0.078470993$ & $0.063317266$ & $0.001127264$ & $-0.000015362$ & $-0.014041825$ \\
 406 & $-0.078622237$ & $0.063461817$ & $0.001121835$ & $-0.000015335$ & $-0.014053920$ \\
 595 & $-0.078756303$ & $0.063591427$ & $0.001118551$ & $-0.000015304$ & $-0.014061628$ \\
 872 & $-0.078842674$ & $0.063674480$ & $0.001115578$ & $-0.000015264$ & $-0.014067879$ \\
1278 & $-0.078927592$ & $0.063758450$ & $0.001114301$ & $-0.000015230$ & $-0.014070071$ \\
1873 & $-0.078982156$ & $0.063811519$ & $0.001113287$ & $-0.000015222$ & $-0.014072572$ \\
2745 & $-0.079013248$ & $0.063842513$ & $0.001112827$ & $-0.000015209$ & $-0.014073117$ \\
4023 & $-0.079034180$ & $0.063862568$ & $0.001112179$ & $-0.000015203$ & $-0.014074637$ \\
5896 & $-0.079045823$ & $0.063874616$ & $0.001111739$ & $-0.000015200$ & $-0.014074667$ \\ \hline
HF$^a$ & $-0.07799$ & $0.06300$ & $0.00132$ & --- & $-0.01367$ \\
MCHF$^b$ & $-0.078935$ & $0.063934$ & $0.001319$ & --- & $-0.013682$ \\
DMC$^c$ & $-0.0790(2)$ & $0.0639(2)$ & $0.001115(8)$ & $-0.000017(2)$ & $-0.0140(3)$ \\ \hline
\multicolumn{6}{|l|}{$^5S_u$} \\ \hline
  88 & $-0.075975991$ & $0.061299742$ & $0.001088726$ & $0.000056199$ & $-0.013531324$ \\
 129 & $-0.076395863$ & $0.061707535$ & $0.001085135$ & $0.000056148$ & $-0.013547045$ \\
 189 & $-0.076664784$ & $0.061964780$ & $0.001082663$ & $0.000056189$ & $-0.013561152$ \\
 277 & $-0.076929320$ & $0.062222229$ & $0.001080459$ & $0.000056205$ & $-0.013570427$ \\
 406 & $-0.077035132$ & $0.062326668$ & $0.001077918$ & $0.000056184$ & $-0.013574362$ \\
 595 & $-0.077161830$ & $0.062450289$ & $0.001077187$ & $0.000056167$ & $-0.013578169$ \\
 872 & $-0.077223369$ & $0.062511108$ & $0.001075924$ & $0.000056169$ & $-0.013580168$ \\
1278 & $-0.077236771$ & $0.062524877$ & $0.001075463$ & $0.000056165$ & $-0.013580265$ \\
1873 & $-0.077289679$ & $0.062575932$ & $0.001074897$ & $0.000056165$ & $-0.013582685$ \\
2745 & $-0.077300396$ & $0.062587016$ & $0.001074547$ & $0.000056165$ & $-0.013582669$ \\
4023 & $-0.077320425$ & $0.062606802$ & $0.001074403$ & $0.000056164$ & $-0.013583055$ \\ \hline
HF$^a$ & $-0.07623$ & $0.06171$ & $0.00127$ & --- & $-0.01325$ \\
MCHF$^b$ & $-0.077202$ & $0.062654$ & $0.001278$ & --- & $-0.013270$ \\ \hline
\multicolumn{6}{l}{$^a$Hartree-Fock \cite{Davidson-CH2},
$^b$multiconfiguration Hartree-Fock \cite{C-MCHF}},
$^c$Quantum Monte Carlo \cite{atoms-Seth}
\end{tabular}
\end{table}

Notice should be taken, that the present result is almost equal to that
published in Ref. \cite{C-MCHF} (33711~cm$^{-1}$), but the latter is
accurate owing to a fortunate cancellation of errors.
The finite nuclear mass effect (about $-9$cm$^{-1}$) was not taken into
account there. The orbit-orbit magnetic interaction energy was also omitted,
resulting in underestimated relativistic correction to the excitation energy
in older calculations \cite{Davidson-CH2,C-MCHF}. This term has
the smallest absolute value among all components of scalar Breit-Pauli
corrections, but is the one with opposite signs for $^3P_g$ and $^5S_u$
states. It contributes nearly 16 cm$^{-1}$ to the excitation energy -- almost
15\% of the total contribution of relativistic corrections, amounting to
108 cm$^{-1}$.

\section{Conclusions}

The optimized ECG lobe functions, projected onto proper representations of
finite point groups, appear to be a powerful tool for studying the
properties of atomic states. Quite surprisingly, they form a
more efficient basis, giving lower variational energies at noticeably
shorter expansions, than the ECGs with preexponential factors, which are
eigenfunctions of $\hat{L}^2$. For the first time, the nonrelativistic
energies of an 6-electron atom were calculated with accuracy better than
20~$\mu$hartree.
Apparent weakness, manifesting oneself in $\langle L^2\rangle$
deviating from exact eigenvalue of this operator, may be utilized for energy
extrapolation, leading to
new estimations of nonrelativistic energies of the lowest triplet
and quintuplet states of the carbon atom.

There are $^1D_g$ and $^1S_g$ states of the carbon atom, stemming from
the same orbital electron configuration as the ground state ($1s^22s^22p^2$),
and with energies lower than that of the $^5S_u$ term. They were not studied
here, due to excessive computational cost of the optimization of thousands
of nonlinear parameters of explicitly correlated basis functions, and they
are postponed to a future work. Realization of the present project lasted
for about two years and engaged varying number of modern CPU cores, reaching 
several hundreds working in parallel on calculation of matrix elements.

Concerning the goal to achieve spectroscopic accuracy of quantum-chemical
calculations for the carbon atom, there is still a long way to go.
The experimental accuracy of the energy difference of the two states
considered here, amounting to 0.018cm$^{-1}$, i.e. 82 nanohartree,
is by two orders of magnitude better than what the present calculations may
offer. The fine structure as well as the contribution of radiative corrections
have to be addressed by future work.

\section{Acknowledgments}

This work was financed by the statutory activity
subsidy from the Polish Ministry of Science and Higher Education for the
Faculty of Chemistry of Wroc{\l}aw University of Science and Technology
(contract number 0401/0121/18). Completion of the project was possible owing to 
extraordinarily friendly policy of Wroc{\l}aw Center for Networking and
Supercomputing (WCSS, http://wcss.pl), where most calculations have been
carried out. Author is grateful for allowing flexible resource allocation,
tailored to the needs of the researcher. Thanks are also due to
dr Pawe{\l} K\k{e}dzierski for maintenance of the computer laboratory,
where small basis sets were optimized before moving to WCSS.

\end{document}